# 86-km optical link with a resolution of $2\times10^{-18}$ for RF frequency transfer.

O. Lopez[1], A. Amy-Klein[1,a], C. Daussy[1], Ch. Chardonnet[1], F. Narbonneau[2], M. Lours[2], and G. Santarelli[2]

[1] Laboratoire de Physique des Lasers, UMR 7538 CNRS, Université Paris 13, 99 av. J.-B. Clément, 93430 Villetaneuse, France

[2] LNE-SYRTE, Observatoire de Paris, 61 Avenue de l'Observatoire, 75014 Paris, France

Abstract : RF frequency transfer over an urban 86 km fibre has been demonstrated with a resolution of $2\times10^{-18}$ at one day measuring time using an optical compensator. This result is obtained with a reference carrier frequency of 1 GHz, and a rapid scrambling of the polarisation state of the input light in order to reduce the sensitivity to the polarisation mode dispersion in the fibre. The limitation due to the fibre chromatic dispersion associated with the laser frequency fluctuations is highlighted and analyzed. A preliminary test of an extended compensated link over 186 km using optical amplifiers gives a resolution below $10^{-17}$ at 1 day.

---

[a] Corrresponding author : fax 33 1 49 40 32 00, Phone : 33 1 49 40 33 79, E-mail : amy@univ-paris13.fr

## 1. Introduction

Besides the current developments of atomic clocks, high resolution frequency distribution and comparison have become a new challenge in the field of frequency metrology. Atomic clocks have improved to a fractional frequency accuracy in the range $10^{-15}$-$10^{-16}$ (see for instance [1-4]) and fractional accuracies in the $10^{-17}$ range or better are under investigation by several groups. The comparison between collocated clocks has been greatly simplified by the recent development of optical synthesizer based on femtosecond laser frequency comb (see for instance [5]). On the other hand the comparison of distant RF or optical clocks is currently using satellite-based techniques and is limited at a level around $10^{-15}$ over one day of measurement. In the context of fundamental physics tests the development of long-distance high resolution frequency transfer is an important issue. From the comparison of atomic or molecular clocks the temporal variation of fundamental constants can be measured providing a test of the equivalence principle [6-7]. The development of such measurements for a large number of clocks with different atoms and molecules will allow discriminating the variations of the different constants and will improve the reliability of measurements in this controversial field. Astrophysics could also benefit from such frequency transfer development especially for large-base interferometers [8-12].

Recent developments gave greater place to optical fibres for high resolution frequency transfer [13]. Progress has been reported in the dissemination of either a RF/microwave [10-12, 14-15] or an optical reference frequency [16-18]. Optical transfer is promising and very attractive in the perspective of optical clocks comparison; alternatively a RF system can be easier to implement depending on the local laboratories facilities and the aimed applications.

We have already demonstrated the distribution of a reference signal in the RF domain at 100 MHz through an optical link of 43 km between two laboratories, LPL and LNE-SYRTE [15, 19]. Since mechanical perturbation and temperature variation along the fibre slightly perturb the propagation delay on the link, a noise compensation set-up was implemented and described in [14-15]. Two types of noise compensator have been developed, both using a round trip phase correction technique. We will design as electronic compensator the one based on the phase conjugation method; it acts on the phase of the reference signal injected in the link [10]. The so-called optical compensator is based on an optomechanical actuator, which acts directly on the optical length of the fibre [8, 11]. With this compensated link, two frequency standards from each laboratory, one in the IR domain ($CO_2$/$OsO_4$ at 30 THz) and the other in the microwave domain (CSO at 12 GHz) have been compared with a resolution of $3 \times 10^{-14}$ at 1s [15].

This paper is describing the improvement of the transfer resolution on the cascaded 86 km link using an upgraded optical compensator. Two major developments have been implemented. To gain on the signal-to-noise ratio, the reference carrier frequency has been increased to 1 GHz. To reduce the sensitivity to polarisation mode dispersion (PMD) in the fibre, the polarisation state of the input light has been scrambled. The further limitation due to the fibre chromatic dispersion is highlighted and analyzed. A preliminary attempt to realize a compensated link over 186 km is also presented.

## 2. RF reference dissemination and noise optical compensation.

The schematic of the optical link and its compensation system is shown in Figure 1. The link connecting LPL and LNE-SYRTE is composed of two parallel 43-km dark optical fibres, which have been obtained by splicing a few sections of SMF 28 single mode fibres from the standard telecommunication network. The 86 km link is obtained by cascading the two twin fibres between LPL and LNE-SYRTE. This way the two ends are collocated at LPL, one designated as Local, and the other as Remote.

The input RF signal at 1 GHz is modulating the intensity of a DFB laser diode which is feeding the optical link. The wavelength is about 1.55 µm and the maximum output optical power is 20 mW. After propagation along the link, the modulation is detected at the other end of the fibre with a fast photodiode, and used to synthesize a coherent RF signal at 900 MHz. This signal exhibits the phase fluctuations of the forward signal at the Remote end; it modulates the intensity of a second laser diode, which is feeding the link in the backward direction. Optical circulators are used at each fibre end to discriminate the forward and backward optical signals. The backward signal is detected at Local end with a second photodiode and exhibits phase fluctuations produced by one round trip. The phase comparison of this backward signal and the input signal is realized at 900 MHz and gives a measurement of the round trip phase fluctuations.

The optical polarizations of both laser diodes are scrambled directly at each fibre end, as discussed below. Two different frequencies, 1 GHz and 900 MHz, are used for the forward and backward RF signals in order to eliminate the parasitic signals due to stray optical reflections on connectors and splicing along the link, which are adding noise to the error signal. The stimulated Brillouin Back Scattering (SBS) is also strongly attenuated by the 100 MHz shift between forward and backward signals, much larger than the SBS gain profile (< 10 MHz).

The error signal is processed by a simple analog loop filter and applied to two delay lines, designed as fast and slow ones, at the Local input of the link in order to cancel out the variations of the propagation delay. The fast delay line is achieved with a 15 m length optical fibre that is wrapped around a cylindrical piezoelectrical transducer (PZT). Fast and small variation corrections are applied on the PZT voltage to stretch the fibre with a correction range of about 15 ps. The slow delay line is made of a 4 km optical fibre wrapped around a copper wheel ; the slow corrections are applied by heating the whole spool with a sensitivity of 150 ps/°C and a total dynamic range of 6 ns. The loop bandwidth is limited by the roundtrip propagation delay (0.88 ms) to a few hundreds Hz. The relative stability of the compensated link is measured by analyzing the phase variation between the 1 GHz signals at both Local and Remote end. The phase signal is filtered with a low-pass filter (3 Hz bandwidth) then digitized and the overlapping Allan deviation of the measured phase samples is deduced.

**3. Results and limitations.**

Figure 2 displays a sample of the phase fluctuation of the compensated link. The propagation delay fluctuation is around 1 ps over a few days. The fractional frequency stability of the link is displayed on Figure 3 (squares): it is $5 \times 10^{-15}$ at 1 second integration time and $2 \times 10^{-18}$ at one day integration time. For comparison, without any corrections, the stability is $3 \times 10^{-14}$ at 1 second with a floor around $10^{-15}$ after 100s (triangles in Figure 3) limited by the diurnal temperature variations. The three orders of magnitude reduction of the long-term frequency fluctuations demonstrates the effectiveness of the compensation system. The stability at one day is 5 times better than previous implementation of the link at 100 MHz on 43 km (circles in figure 3) [15]. Moreover the average frequency offset between reference signal and transferred signal over the link is zero within the error bars. The better performance of the present system is directly related with the increased of the RF carrier frequency from 100 MHz to 1 GHz (900 MHz for the backward signal). The signal-to-noise ratio of the error signal driving the compensator is increased about a factor 9: fibre phase fluctuations are 9 times higher while detection noise is approximately the same.

The performance of this round trip compensation technique is limited by the parasitic noise on the error signal. Figure 4 shows the phase noise spectral density at 1 GHz of the compensator in closed loop with the 86 km link (upper trace) and with an equivalent optical attenuator instead of the link (lower trace). This last trace displays the noise of the compensation system, which is at the level of -120 dB at 1 Hz. It corresponds to an Allan deviation of $10^{-15}$ at 1 s integration time, which is the aimed stability for the compensated 86-km link. However the optical link degrades the noise by nearly 15 dB, leading to a noise of -105 dB at 1 Hz. The bump around 300 Hz is due to the bandwidth of the control loop.

This parasitic excess phase noise originates from the fact that the round trip phase fluctuations do not exactly correspond to twice the forward phase fluctuations due to an asymmetry of the phase noise in forward and backward propagation directions along the link. Due to the chromatic and polarization mode dispersions of the fibre the propagation delay is sensitive to the characteristics of the two optical signals injected in each direction at each end of the link. The two laser diodes have uncorrelated optical frequency fluctuations and polarizations state. We show in the next paragraph how the asymmetry due to the polarization mode dispersion (PMD) can be strongly reduced by scrambling the light polarization state at each end. Thus, in our set-up, the short-term frequency stability is limited by the chromatic dispersion and laser frequency fluctuations. The long-term frequency stability is restricted by the RF electronic components temperature sensitivity in the compensator.

**4. Polarization effects.**

The propagation characteristics along the fibre depend on the polarization state through the polarization mode dispersion, PMD (see for instance [20]). The PMD is induced by the birefringence of the optical fibre due to the asymmetry of the core and is varying in time because of the mechanical stress induced by vibrations or by temperature variations. Moreover the directions of both fast and slow axes of the link move randomly in time.

Due to PMD, the propagation delay along the link fluctuates. Since two independent laser diodes with different wavelengths and polarisation states are used at each fibre end, these propagation delay fluctuations are uncorrelated for the forward and backward signals along the link. This lack of symmetry of the propagation delay fluctuations degrades the effectiveness of the noise compensator, which can only correct the symmetrical phase fluctuations, since the error signal is the roundtrip phase fluctuations. The main effect is the degradation of the long term stability.

The polarisation related effects are enhanced by the optical compensator itself. Both delay lines induce a parasitic change of the polarization state of the signal injected in the 86-km link and thus introduce an unwanted variation in the propagation delay along the link. This variation is caused by the stretching of the correction fibres, and can be much faster than the typical slow variation of PMD.

As mentioned above, polarization effects can be significantly reduced by scrambling the polarization directly at each laser diode output, at a frequency higher than the cut off frequency of the stabilization loop. The polarization scrambling is obtained with a commercial three axes piezoelectric polarization scrambler. Each axis is excited with a low voltage sinusoidal signal tuned to a piezoelectric resonance. Among the possible resonance we chose the non-harmonically related ones (around 60, 100, and 130 kHz respectively). This fast polarisation modulation explores all the polarization states much faster than the propagation delay along the link [21]. In this case the DGD is averaged and the PMD effects are minimized.

In order to evaluate the limitations due to the PMD or the delay line induced polarisation effects, the following measurements have been performed. The phase fluctuations between the local and remote ends have been first recorded without any scramblers and the corresponding Allan deviation is plot as triangles on Figure 5. In that case, the fast delay line is not used. In fact, we may notice that the delay lines are set directly at the input end of the link. They affect the polarisation from the input of the fibre, and their effect on the propagation delay fluctuations is then enhanced by the PMD. With the PZT fibre stretcher this effect is high enough to perturb the correction loop and prevent its use. Thus for time scale below 100s there is no correction and the fractional frequency stability is that of the free link. For time scale above 100s the stability is degraded by the polarisations effects about one order of magnitude. The bump between $10^3$ and $10^4$ s is due to the slow thermal compensator, which is also changing the polarisation. Since this Allan deviation is not typical from PMD effects, a second experiment has been performed with only one scrambler at the Local end (that is on the forward way) and no scrambler at the Remote end on the backward way. In that case, both delay lines are used and the polarisations effects due to the delay lines and PMD are minimized on the forward propagation delay. On the backward propagation way however, PMD effects are still effective but they are not enhanced by the delay line induced polarisation effects since the delay lines are at the output end of the backward way. With this scheme we aim to test the PMD of the 86-km fibre link alone. The phase fluctuations are displayed on Figure 2 together with the phase fluctuations of the compensated link with the two scramblers. Let us remind that the compensation system does not distinguish the forward and backward fluctuations and the delay lines are correcting on each way half the total round trip fluctuations. Since the PMD affects solely the backward phase fluctuations, only half of the PMD phase fluctuations can be corrected, and the phase data in closed loop still exhibits half the PMD phase fluctuations on the backward way. It amounts around 2 ps, and thus the effects of PMD on one way is about 4 ps. The Allan deviation plot of the phase data of Figure 2 is displayed as circles on Figure 5 and compared to the Allan deviation plot with two scramblers (squares). The PMD does not affect the short-term stability but diurnal temperature variations of the PMD degrade the long-term frequency stability to a few $10^{-17}$.

Finally, when scrambling the polarization, the stability (squares in Figures 3 or 5) reaches a value compatible with the laser dispersion noise level (see section 5 below), which demonstrates that the polarization effects are strongly cancelled out.

## 5. Chromatic dispersion effects.

The second source of parasitic noise limiting the compensation performance arises from the chromatic dispersion of the fibre associated with the frequency fluctuations of the laser diodes. An additional phase noise, that we will call dispersion noise, is induced along the fibre, and is not correlated in the forward and backward signals. This asymmetrical additional phase noise can not be corrected by our compensation system.

The laser current modulation induces both an amplitude and a frequency modulation (chirp), and the spectrum of both laser diodes is constituted by a carrier at the optical frequency and sidebands separated by the reference RF frequency $\frac{\Omega}{2\pi}$ (= 0.9 or 1 GHz) [22].

The detected signal at the RF frequency results from the beat notes between adjacent lines, which propagate at different velocities due to the dispersion. At first order, the propagation time of two adjacent lines differs from the quantity $\Delta t_d = -D L \left( \frac{\lambda_0 \, \Omega}{\omega_0} \right)$ where D=17

ps/km.nm is the dispersion of the fibre, L the fibre length (in km), $\lambda_0$ the laser wavelength (in nm) and $\frac{\omega_0}{2\pi}$ the laser frequency. In our case at 1 GHz, $\Delta t_d = -12.3 \, ps$ for the 90-km fibre (including the link and the correction fibre spool). The phase detected at each end of the fibre at 1GHz (respectively 900 MHz) is the contribution of two terms: $\Delta\varphi = \Omega.t_0 + \omega_o.\Delta t_d$ where $t_0$ is the propagation time of the laser carrier along the fibre (see detailed calculation in Annex). The fluctuation of $t_0$ in the first term gives the fibre noise that we want to correct. The second term gives the so-called dispersion noise resulting from the conversion of the laser frequency noise by the fibre chromatic dispersion. The resulting dispersion noise after one round trip is $(\delta\omega_{LD1} + \delta\omega_{LD2}).\Delta t_d$ where $\delta\omega_{LD1}$ and $\delta\omega_{LD2}$ are the uncorrelated frequency fluctuations of each laser diode. The dispersion noise degrades the system phase noise level of around 15 dB (see Figure 4), leading to a noise of -105 dB at 1 Hz.

At the remote end, the remaining phase noise is $\delta\phi_d = \frac{(\delta\omega_{LD1} + \delta\omega_{LD2})}{2}.\Delta t_d$ in closed loop. This additional phase noise can be evaluated by measuring the beat-note of the two laser diodes with a counter (Allan deviation ≈ 250 kHz at 1s). The laser frequency fluctuations $\frac{\delta\omega_o}{2\pi} = \frac{\delta\omega_{LD1} + \delta\omega_{LD2}}{2\pi}$ is then converted into the above RF phase noise $\delta\phi_d = \frac{\delta\omega_o}{2}.\Delta t_d$. The Allan deviation floor (star plot) due to this effect is displayed in Figure 3 and is very close to the Allan deviation of the compensated link. The small difference partly arises from the additional thermal fluctuations induced by the compensator on the laser diodes. This demonstrates that the laser frequency noise associated with the dispersion of the fibre is presently the main limitation for the resolution of the compensation link. It is worth noting that this contribution to the noise scales linearly with the modulation frequency through $\Delta t_d$ and, thus, cannot be reduced by increasing this value.

This limitation could be overcome with a stabilisation of the laser diodes frequencies. Extensive efforts had already been done to reduce the laser frequency fluctuations with a better control of the diode temperature and bias current. An active stabilisation is more efficient although not straightforward to implement, especially at long-term. Laser frequency fluctuation of the order of 1 kHz at 1 s would lead to a negligible noise contribution (at the level of $10^{-17}$). Finally a more simple solution would consist in adding a negative dispersion fibre of adequate length. This would compensate for the link chromatic dispersion, even if losses will be inserted.

**6. Longer link: the problem of the attenuation.**

The above results demonstrate that an RF optical link can be used for metrological purposes to disseminate a RF reference signal between laboratories in the same area. We present a preliminary evaluation of this technique for longer distances, which has been tested by adding fibre spools to the existing 86-km link.

For a longer link, the performance of the noise compensator is still limited by the dispersion noise, which increases proportionally to the fibre length. Moreover the signal-to-noise ratio is further reduced by the attenuation of the fibre spool, which yields a reduction of 40dB on the RF signal for each 100 km of additional fibre. This attenuation can not be compensated by increasing the laser diodes power, since the power in each sideband is already near the threshold of SBS. To overcome this limitation we use Erbium Doped Fibre

Amplifiers (EDFA). A preliminary test was done with a 186-km link composed of the 86-km link and 100 km of fibre spools. Two standard telecom unidirectional amplifiers were connected with two optical circulators in order to obtain a full bidirectional configuration. This composite bidirectional EDFA was inserted at half distance of the link and the amplification level was adjusted to compensate the additional optical losses. For such an amplification level (around 20 dB), the EDFA's excess phase noise is equivalent to a few $10^{-15}$ frequency stability at 1s, which is negligible compared to the chromatic dispersion noise.

Figure 6 shows preliminary results of the relative frequency stability of this 186 km link for two different set-ups, one using the optical compensator and the other an electronic compensator. For this alternative compensator described in [14], the correction is applied directly on the RF source signal. The two systems give about the same results. At short-term, the Allan deviation for the optical compensator is $2\times10^{-14}$ at 1 s integration time, whereas the limit due to the laser dispersion is $1\times10^{-14}$ at 1s. This small difference could be attributed to any optical or electronic components which have not been carefully optimised for these measurements. The long term frequency stability is better than $10^{-17}$ at one day integration time.

This demonstrates the first realization of a RF frequency transfer below $10^{-17}$ level on distances above 100 km. The resolution is very close to the limit due to the dispersion noise, which demonstrates that EDFA's can be used to compensate for the attenuation. For longer distances the stability is degraded because the fibre phase fluctuations and the loop bandwidth are respectively increasing and reducing with the fibre length. For a 1000 km link around 10 EDFA's should be used to recover the signal, the control bandwidth is reduced to around 10 Hz, and the fibre noise could not be corrected by more than a factor of 10 at 1 s. However the present set-up demonstrates an elementary component which could act as a repeater for frequency dissemination over a multi-section optical link. By cascading a few such systems the stability of around $10^{-17}$ at one day could be maintained over distances up to 1000 km.

## 7. Conclusion.

We have demonstrated a 86-km compensated link from an urban network with a frequency stability of $5\times10^{-15}$ at one second and $2\times10^{-18}$ at one day integration time. These results were obtained after the limiting effects were analyzed and two different AM frequencies used for each propagation direction and light polarization state scrambled. The dispersion of the fibre, associated with the laser frequency fluctuations, gives the present limit of our compensation set-up. This limitation can be overcome by the laser frequency stabilisation or by the cancelling of the chromatic dispersion with a negative dispersion fibre. In that case, an Allan deviation of $10^{-15}$ at 1 s is achievable, limited by the electronic noise of the compensator. For a longer link, the signal attenuation becomes critical. For 186-km, a stability below $10^{-17}$ at one day integration time has been measured using EDFA's and the short-term is limited by the fibre chromatic dispersion. This demonstrates a quite simple repeater for multi-section frequency dissemination over very long distances.

The compensator could be further improved by using a microwave frequency signal around 10 GHz instead of the present 1 GHz signal. The signal-to-noise ratio of the error signal is indeed proportional to the reference frequency. Thus the distance between repeaters for multi-section frequency transfer could be chosen with an additional tolerance of 50 km. The degradation of the error signal due to the EDFA's and the electronic noise is also much less critical. However the set-up and compensator should be adapted to microwave modulation techniques and the stability improvement is not straightforward.

The present results give a new opportunity to compare distant frequency standards. Our compensation system constitutes a robust repeater with 100 % duty cycle and a resolution below $10^{-17}$ at 1 day; it could be used for multi-section frequency transfer up to 1000 km distances. Further developments should render optical links a real alternative to satellite link.

Acknowledgments
This work was supported by Ministère de la Recherche and European Space Agency/ESOC.

**Appendix :**

The laser field is given by $E = E_0\sqrt{1+m_i\cos\Omega t}\ e^{j(\omega_0 t + m\sin\Omega t)}$ where $m_i$ and $m$ are respectively the amplitude and frequency modulation indexes. In our case, $m_i=0.7$ and $m=15$, due to the chirp coefficient of the laser diode estimated to $\eta=375$ MHz/mA [20, 22].

The field can be developed in Fourier series at the modulation frequency $\dfrac{\Omega}{2\pi}$. We note $J_n = J_n(m)$ the coefficients of the Fourier serie of the phase, and $M_n$ the coefficients of the Fourier serie of the amplitude $\sqrt{1+m_i\cos\Omega t} = \sum_{n=0}^{\infty} M_n \cos n\Omega t$.

One obtains an asymmetrical spectrum [22] :

$$E = E_0 e^{j\omega_0 t}\left(L_0 + \sum_{n=1}^{\infty}\left(L_{n+}e^{jn\Omega t} + L_{n-}e^{-jn\Omega t}\right)\right)$$

with
$$\begin{cases} L_0 = \sum_{a=0}^{\infty} M_{2a} J_{2a} \\ L_{n+} = \dfrac{1}{2}(M_0 J_n + J_0 M_n) + \dfrac{1}{2}\sum_{a=1}^{\infty} J_a\left(M_{|n-a|} + (-1)^a M_{n+a}\right) \\ L_{n-} = \dfrac{1}{2}\left((-1)^n M_0 J_n + J_0 M_n\right) + \dfrac{1}{2}\sum_{a=1}^{\infty} J_a\left(M_{n+a} + (-1)^a M_{|n-a|}\right) \end{cases}.$$

After propagation along the fibre, the Fourier components are dephased by the quantities $\varphi_0$ (for the fundamental) and $\varphi_{n+}$ and $\varphi_{n-}$ (for the $+n\Omega$ and $-n\Omega$ components respectively).

At low order (in term of chromatic dispersion), we have :
$$\begin{cases} \varphi_0 - \varphi_{1+} = -\omega_0\Delta t_d - \Omega\Delta t_d - \Omega t_0 \\ \varphi_0 - \varphi_{1-} = \omega_0\Delta t_d - \Omega\Delta t_d + \Omega t_0 \\ \varphi_{1+} - \varphi_{2+} = -\omega_0\Delta t_d - 3\Omega\Delta t_d - \Omega t_0 \\ \varphi_{1-} - \varphi_{2-} = \omega_0\Delta t_d - 3\Omega\Delta t_d + \Omega t_0 \\ \ldots \end{cases}.$$

The intensity is detected and filtered at $\dfrac{\Omega}{2\pi}$. One obtains:

$$S = 2\sum_{n=0}^{\infty}\left(L_{n+}L_{(n+1)+} + L_{n-}L_{(n+1)-}\right)\cos\left((2n+1)\Omega\Delta t_d\right)\cos\left(\Omega t - \omega_0\Delta t_d - \Omega t_0\right)$$
$$+ 2\sum_{n=0}^{\infty}\left(L_{n+}L_{(n+1)+} - L_{n-}L_{(n+1)-}\right)\sin\left((2n+1)\Omega\Delta t_d\right)\sin\left(\Omega t - \omega_0\Delta t_d - \Omega t_0\right)$$

The term in quadrature is resulting from the frequency modulation of the laser, which induces a non-symmetrical spectrum. Note also that the amplitude is depending from the differential time delay $\Delta t_d$. The amplitude is thus modulated with the fibre length.

**Figure captions**

Figure 1: Block diagram of the 86 km fibre link and the optical compensation system. Both local and remote functions are located in the same place. DFB LD : distributed feedback laser diode.

Figure 2: Propagation delay fluctuation of the compensated link (a) with 2 scramblers and (b) with only one scrambler (see section 4); it is obtained from the differential phase fluctuation of the 1 GHz signals at local and remote ends.

Figure 3: Fractional frequency stability of the free 86-km link (triangles), the compensated link at 100 MHz (circles) and 1 GHz (squares) and dispersion noise due to the laser diodes frequency fluctuation (stars).

Figure 4: Phase noise spectral density of the optical compensation system at 1GHz in closed loop a) with the 86-km optical fibre b) with optical attenuators equivalent to the fibre attenuation.

Figure 5: Polarisation effects: fractional frequency stability of the 86-km compensated link without scramblers (and no fast delay line) (triangles), with only one scrambler at input end (circles), and with two scramblers (squares).

Figure 6: Preliminary fractional frequency stability of the 186-km link. Triangles: electronic compensator; squares: optical compensator.

Figure 1

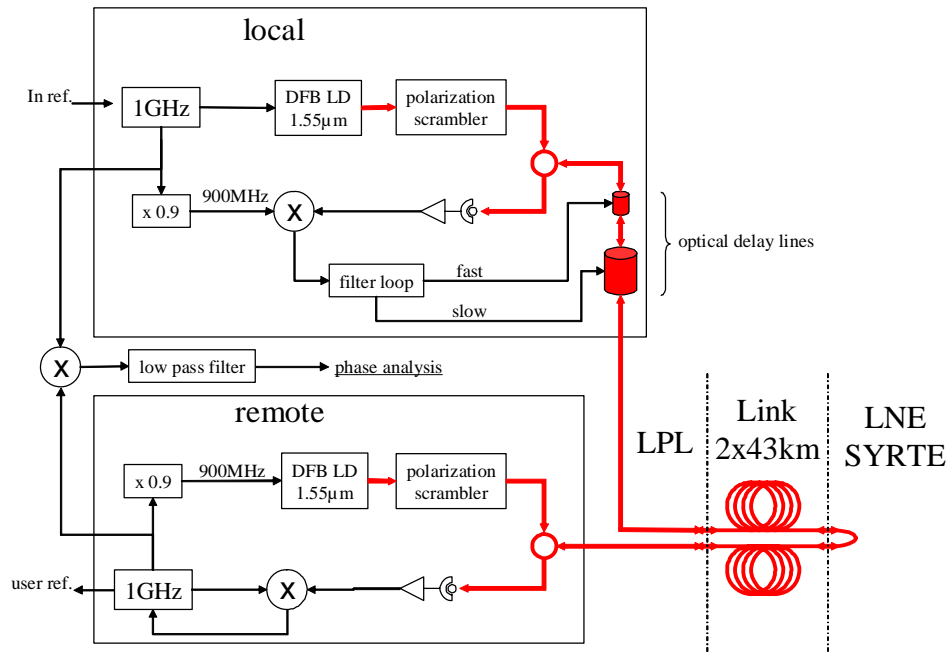

Figure 2

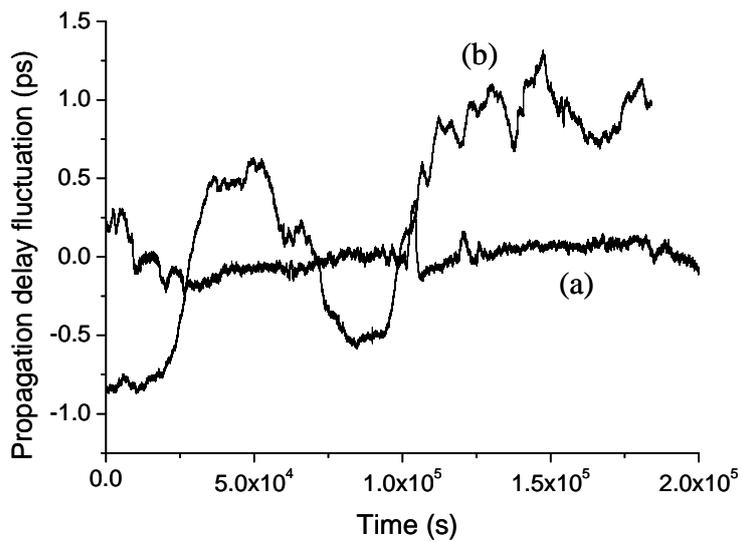

Figure 3

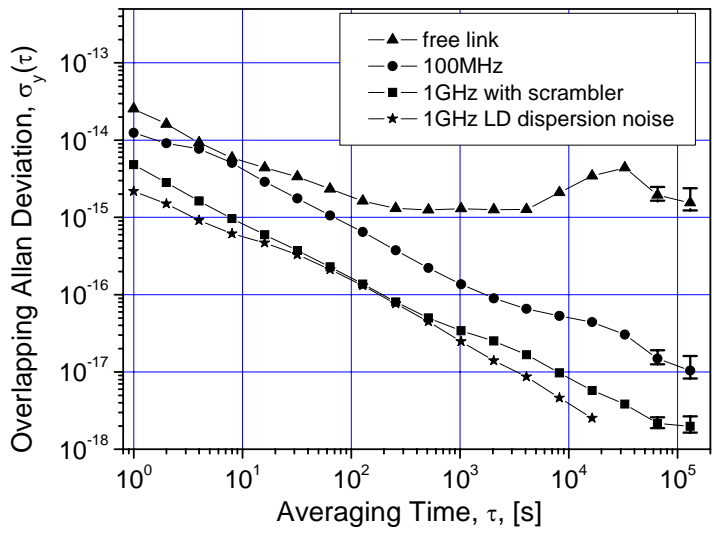

Figure 4

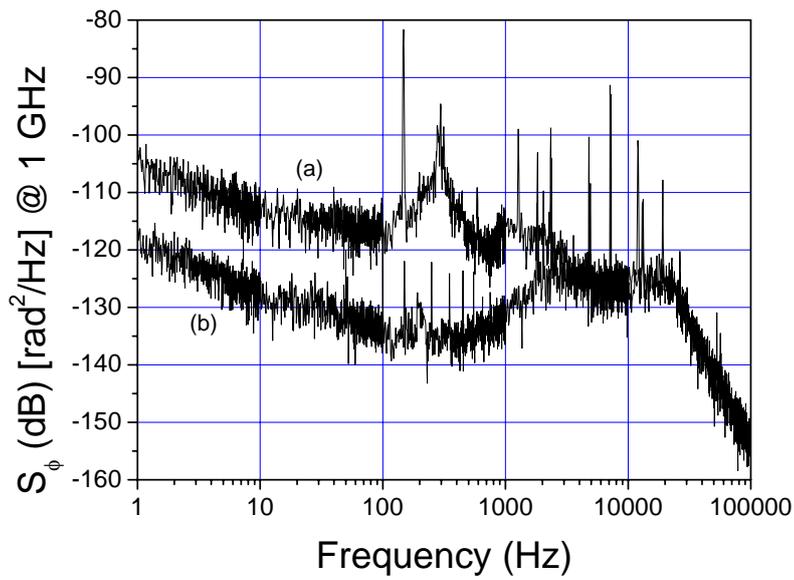

Figure 5

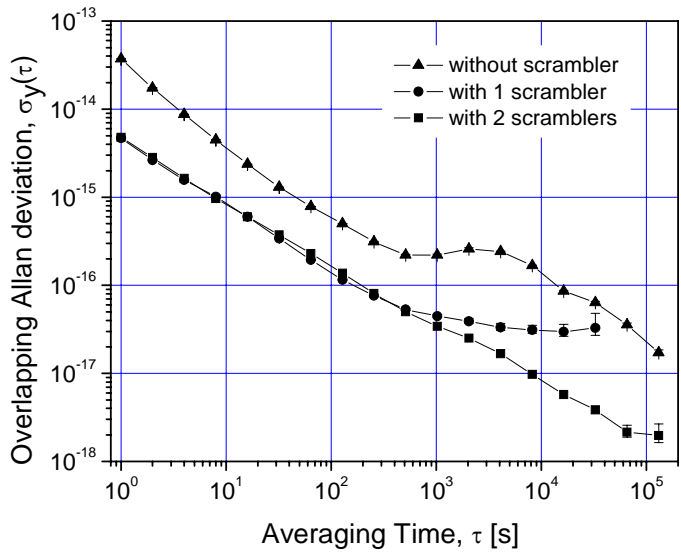

Figure 6

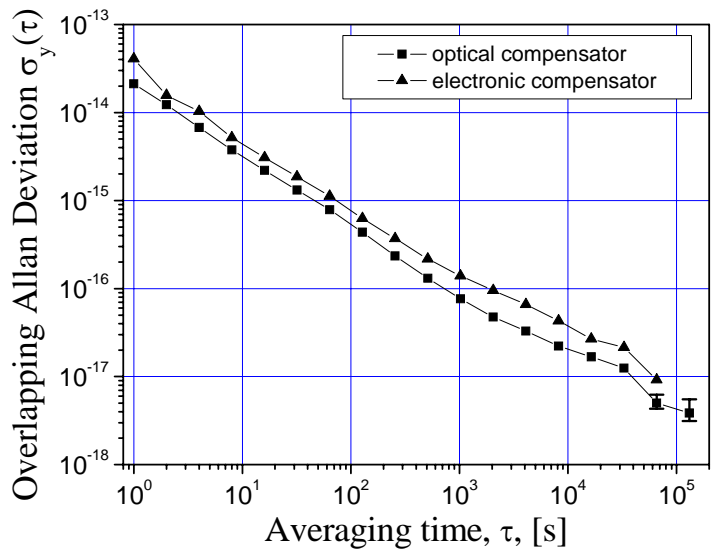